\documentclass[12pt]{article}

\usepackage{setspace}
\usepackage[square,numbers,compress]{natbib}
\usepackage[utf8]{inputenc}
\usepackage{amsfonts}
\usepackage{amssymb}
\usepackage{multirow}
\usepackage{amsmath}
\usepackage{nth}
\usepackage{titling}
\usepackage{mathrsfs}
\usepackage{lscape}
\usepackage[margin=1.in]{geometry}
\usepackage{caption}
\usepackage{graphicx}
\usepackage{subcaption}
\usepackage{setspace}
\usepackage{multibib}
\usepackage{lipsum}
\usepackage{tcolorbox}
\usepackage{xcolor}
\usepackage[section]{placeins}
\usepackage{chngcntr}
\usepackage{newfloat}
\usepackage{booktabs}
\usepackage{tabularx}
\usepackage{ltablex}
\usepackage{multicol}
\usepackage{authblk}
\usepackage{lineno}
\usepackage{har2nat}
\usepackage{setspace}
\usepackage[onelanguage,ruled,vlined,linesnumbered,resetcount]{algorithm2e}
\usepackage[pdftex,hidelinks]{hyperref}
 
\usepackage[section]{placeins}
\usepackage{authblk}

\title{A Method for Emerging Empirical Age Structures in Agent-Based Models with Exogenous Survival Probabilities}

\author[1,*]{Kathyrn R. Fair}
\author[1]{Omar A. Guerrero}

\affil[1]{The Alan Turing Institute, London}
\affil[*]{kfair@turing.ac.uk}

\date{}

\begin{document}

\maketitle

\begin{abstract}
For many applications of agent-based models (ABMs), an agent's age influences important decisions (e.g. their contribution to/withdrawal from pension funds, their level of risk aversion in decision-making, etc.) and outcomes in their life cycle (e.g. their susceptibility to disease).
These considerations make it crucial to accurately capture the age distribution of the population being considered.
Often, empirical survival probabilities cannot be used in ABMs to generate the observed age structure due to discrepancies between samples or models (between the ABM and the survival statistical model used to produce empirical rates).
In these cases, imputing empirical survival probabilities will not generate the observed age structure of the population, and assumptions such as exogenous agent inflows are necessary (but not necessarily empirically valid).
In this paper, we propose a method  that allows for the preservation of agent age-structure without the exogenous influx of agents, even when only a subset of the population is being modelled. 
We demonstrate the flexibility and accuracy of our methodology by performing simulations of several real-world age distributions. 
This method is a useful tool for those developing ABMs across a broad range of applications.
\end{abstract}


\section{Introduction}

There exists a broad class of agent-based models (ABMs) in which (1) agents become older with time, (2) their survival probability is exogenous and dependent on age, (3) the population size is assumed constant, (4) the dynamics reach a steady state, and, sometimes, (5) their population does not cover the entire age spectrum.
Examples of such models can be found in labour studies \cite{axtell_coordination_2012,goudet_worksim_2017,fair2023emerging} (see \cite{neugart_agentbased_2018} for a comprehensive review), demography \cite{billari_agentbased_2003,diaz_agent_2010}, and microsimulation \cite{cai_accounting_2006,leombruni_laborsim_2006,reznik_longevityrelated_2019,borella_microsimulation_2010}.
Frequently, these applications are not concerned with preserving the empirical age distribution of the (sub)population, but with accounting for empirical survival probabilities.
In some cases, preserving the real age distribution is a priority, for example, when focusing on the adult labour force with the aim of studying retirement- and life-cycle-related problems.
Often, and especially when working with sub-populations, ABMs fail at preserving the empirical age distributions if supplied with official survival probabilities.
This problem arises from a mismatch between the assumed statistical model used in survival analysis and the one underlying the ABM, as well as in the populations covered.

A naive approach to preserve the age structure of a sub-population is to re-sample the empirical age distribution whenever an agent dies.
While this indeed solves the preservation problem, it is not ideal in applications that study the life cycle of the agent, for example, when considering career progression.
Two other popular remedies are imposing coarse age bins and exogenous migration inflows.
The former destroys information on age structure that could be useful to address problems related to highly stratified populations.
The latter (migration flows) may not be empirically valid (or observed), but rather an artefact to generate the empirical age distributions.
An alternative framework aiming at preserving the first two moments of the age distribution has been proposed by \cite{schindler_importance_2012}.
This approach also has limits as nuanced structures in the shape of the distribution may be key for applications related to the life cycle of the agent.

In this paper, we develop a method to estimate age-specific survival probabilities that generate arbitrary age structure in an ABM that meets the four conditions mentioned above.\footnote{Note that a large part of the ABM demographic literature is concerned about population growth and changes in the age distribution.
This method does not address those problems, but only the subset of models that meet the aforementioned characteristics.
Likewise, public health scholars may be interested in endogenous survival probabilities (e.g., as a result of transmissible diseases).
Our approach is not designed to be used with such models as the survival probabilities need to be exogenous.}
Our method consists of a stochastic process of an ageing population and a solution for arbitrary steady-state age distributions.
The solution yields a vector with age-specific survival probabilities that guarantees the emergence of the desired age structure (but that may differ from official survival rates).
These survival probabilities can then be imputed in an ABM that considers an ageing process with exogenous survival rates.

\section{Method}

We begin by describing the stochastic process that generates the desired steady-state age distribution.
Then, we outline an extended version of this model that provides more flexibility in terms of the age distribution shapes, and the method used to parameterise it.

\subsection{Model 1: simple stochastic process}

Let us explain the method through the example of a labour market model focusing on adult age groups, from 18 years old to 90 or more years old.
Consider the distribution of the agents across these age groups as $N_{18}, N_{19}, \dots, N_{90+}$.
For each of these groups, their survival probabilities are $p_{18}, p_{19}, \dots, p_{90+}$.
The stochastic process consists of a population with $N$ agents, each one ageing by one year each period and surviving with the probability of their age group.
Every time an agent dies, it is replaced by a new agent that is 18 years old.
Thus, the model assumes a constant population, and ensures that all agents have gone through the same life cycle process.

We can describe the dynamics of this system through the changes of the age groups.
Because of the specifics of the agent-replacement rules, we need to write three different evolution equations: one for the 18-years old group, another for the 90+ group, and another for all those groups in between these two.
Let us begin with the 18-years old one, which is given by

\begin{equation}
    N_{18,t+1} = N_{18,t} - p_{18}N_{18,t} + (1-p_{19})N_{19,t} + (1-p_{20})N_{20,t} + \dots + (1-p_{90+})N_{90+,t},
\end{equation}
which can be read as the number of 18-year old from the previous period minus those who survive and move on to the next group, plus all the agents from the other groups who died (because they are replaced by agents in this group).

The dynamics of the second type can be described as

\begin{equation}
    N_{90+,t+1} =  N_{90+,t} - (1-p_{90+})N_{90+,t} + p_{89}N_{89+,t},
\end{equation}
which consists of the population with 90 or more years from the previous period, minus those who died, plus those from the previous group (with 89 years old) who survived.

Finally, the third type consists of any age group $18 < i < 90$, and can be described by

\begin{equation}
    N_{i,t+1} = N_{i,t} - (1-p_i)N_{i,t} - p_iN_{i,t} + p_{i-1}N_{i-1,t},
\end{equation}
which consists of the surviving population from the previous age group (since all the agents in this group leave it regardless if they die or survive).

\begin{figure}[ht]
	\centering
	\includegraphics[width=0.8\textwidth]{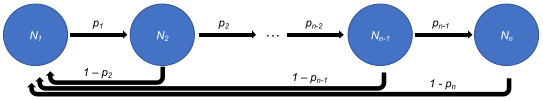}
	\caption{\textbf{Schematic of model 1.} Circles indicate age groups, arrows indicate probability of movement between groups.}
	\label{figure:model}
\end{figure}

This stochastic process leads to a steady state.
In such state, it is safe to assume $N_{i,t+1} = N_{i,t}$ for any age group.
Therefore, by simplifying the three equations and re-arranging them, we can write down the linear system

\begin{equation}
    \begin{split}
        -p_{18}N_{18,t} + (1-p_{19})N_{19,t} + \dots + (1-p_{90+})N_{90+,t} = 0\\
        \vdots\\
        0 + \dots + p_{i-1}N_{i-1,t} - N_{i,t} + \dots + 0 = 0\\
        \vdots\\
        0 + \dots + p_{89}N_{89+,t} - (1-p_{90+})N_{90+,t} = 0
    \end{split},
\end{equation}
or in matrix form for a generic structure with $n$ age groups
\begin{equation}
    \begin{bmatrix} 
    -p_1 & (1-p_2) & (1-p_3) & (1-p_4) & (1-p_5) & \dots & (1-p_n) \\ 
    p_1 & -1 & 0 & 0 & 0 & \dots & 0 \\ 
    0 & p_2 & -1 & 0 & 0 & \dots & 0 \\
    \vdots & \vdots & \vdots & \vdots & \vdots & \vdots & \vdots \\ 
    0 & \dots & p_{i-1} & -1 & 0 & \dots & 0 \\ 
    \vdots & \vdots & \vdots & \vdots & \vdots & \vdots & \vdots \\ 
    0 & 0 & 0 & 0 & \dots & p_{n-1} & -(1-p_n) \\ 
    \end{bmatrix}
    \begin{bmatrix} 
    N_1 \\ 
    N_2 \\ 
    N_3 \\
    \vdots \\
    N_i \\
    \vdots\\
    N_n \\
    \end{bmatrix} =
    \begin{bmatrix} 
    0 \\ 
    0 \\ 
    0 \\
    \vdots \\
    0 \\
    \vdots\\
    0 \\
    \end{bmatrix},
\end{equation}
as described in \autoref{figure:model}.

Thus, in general, for some system with $n$ age groups, we obtain 
\begin{equation}
    \begin{split}
        p_{i-1}=  N_{i,t}/N_{i-1,t} \  \forall \ i \in \{2, \dots, n-1\}\\
        p_{n-1}= (1-p_{n})N_{n,t}/N_{n-1,t} \\
    \end{split},
\end{equation}
with $p_{n}$ as a free parameter, as this specification of the $p$-values will satisfy
\begin{equation}
        -p_{1}N_{1,t} + (1-p_{2})N_{2,t} + \dots +  (1-p_{n})N_{n,t} = 0
\end{equation}

If we add the constraint that $p_{i} \in [0,1] \ \forall \ i \in \{1, \dots, n-2\}$ then, assuming $N_{i,t} > 0$, we obtain 
\begin{equation}
    \begin{split}
    0 \leq N_{i+1,t} \leq N_{i,t} \ \forall \ i \in \{1, \dots, n-2\}
    \end{split}
\end{equation}
and, again assuming non-zero population sizes (i.e. $N_{n,t}, N_{n-1,t} > 0$), to satisfy the condition that $p_{n-1} \in [0,1]$ we need to choose our free parameter $p_n$ such that
\begin{equation}
    \begin{split}
    1 - N_{n-1,t}/N_{n,t} \leq p_{n} \leq 1
    \end{split}
\end{equation}

This condition means that, for this model specification, the types of age distributions that will emerge have a structure where the number of individuals in a group decreases monotonically with age; something common in many countries, at least approximately.
Next, let us generalise the stochastic process to allow for the emergence of age distributions (which can also be empirically observed) where this monotonic pattern does not hold.

\subsection{Model 2: process with activation rates}

Assume that, in addition to the previous model, there is an activation process (see \cite{axtell_effects_2000} for a review of different activation regimes).
Agents are subjected to the ageing process only if they are active.
Conceptually, this assumption may come across as a strange one.
However, as we show ahead, it allows more flexibility in the types of age distributions that can emerge from the stochastic process.
Thus, if introducing this assumption does not affect the model's interpretation and results in a substantial way, it could be useful to capture other observed age structures in an ABM.\footnote{This approach provides an alternative to models that use broad age groups and exogenous migration flows \cite{brennan2020introducing}, creating flexibility for agents to remain within an age group for more/less time than is spanned by the age group.}

\begin{figure}[ht]
	\centering
	\includegraphics[width=0.8\textwidth]{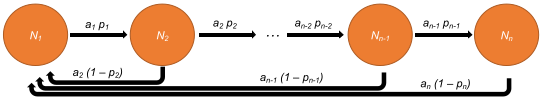}
	\caption{\textbf{Schematic of model 2 (model with activation rates).} Circles indicate age groups, arrows indicate probability of movement between groups.}
	\label{figure:model_activation}
\end{figure}

The activation process in model 2 is governed by a vector of activation rates $\alpha_i, \dots, \alpha_n$.
Parameter $\alpha_i$ determines the activation probability when the agent belongs to age group $i$.
This model is described in \autoref{figure:model_activation}.

Let us focus on the equation describing the intermediate groups, which was given by

\begin{equation}
    N_{i,t+1} = N_{i,t} - (1-p_i)N_{i,t} - p_iN_{i,t} + p_{i-1}N_{i-1,t}
\end{equation}
in model 1, and yields in the steady state

\begin{equation}
    p_{i-1} = N_{i,t}/N_{i-1,t}  \  \forall \ i \in \{2, \dots, n-1\}.
\end{equation}

With the addition of an activation regime, the evolution of intermediate groups is

\begin{equation}
    N_{i,t+1} = N_{i,t} - \alpha_{i}(1-p_i)N_{i,t} - \alpha_{i}p_iN_{i,t} + \alpha_{i-1}p_{i-1}N_{i-1,t}
\end{equation}
which, in the steady state, yields

\begin{equation}
    p_{i-1} = \frac{\alpha_{i}N_{i,t}}{\alpha_{i-1}N_{i-1,t}} \  \forall \ i \in \{2, \dots, n-1\},
\end{equation}
and which allows more flexibility in the $N_{i,t}/N_{i-1,t}$ relationship through the activation rates $\alpha_{i}$ and $\alpha_{i-1}$. 
We also obtain 
\begin{equation}
    p_{n-1} = \frac{\alpha_n (1-p_n)}{\alpha_{n-1}} \frac{N_{n,t}}{N_{n-1,t}}.
\end{equation}

Similarly to the model without activation rates, $p_n$ is left as a free parameter, and we require that $p_i, \alpha_ i \in [0,1] \ \forall \ i \in \{ 1, \dots, n\}$.

\subsection{Parameterisation}

The parameterisation method depends on which of the two models is being used to generate the age distribution.
For the model without activation rates (model 1), the user simply chooses a value for the free parameter ($p_n$) and uses this to calculate the value of $p_{n-1}$, with all $p_i$ for  $\ i \in \{1, \dots, n-2\}$ determined wholly by the observed age distribution.
When using model 2 -- the model with activation rates ($\alpha_i$ for $i \in \{1, \dots, n \}$) -- the process of determining suitable parameter values becomes more complicated as there is no general solution for $\alpha_i$.
Here, we use the heuristic optimisation method of differential evolution to determine suitable parameter values.
In many cases, the optimisation algorithm is able to quickly (both in terms of number of iterations, and in terms of total computational time) determine parameter values that allow the model to replicate the desired age distribution with a high degree of accuracy.
However, for some age distributions, even the addition of activation rates will not provide sufficient flexibility to admit parameter values allowing the model to produce a reasonably accurate age distribution.
In these cases, we suggest the curve-fitting procedure explained below.

\subsection{Fitting distributions}

There exist cases in which the structure of an observed age distribution cannot be generated by either of the two stochastic processes.
In these cases, we suggest modifying the empirical distribution by changing it into a distribution that fulfils the conditions established by the first stochastic process (the one without activation rates).
We can determine the modified distribution by fitting a condition-compliant function to the empirical data to create a modified age distribution.
Determining whether this is a reasonable course of action depends on how different the fitted distribution looks with respect to the original one.
This can be determined by calculating the Wasserstein metric, indicating the minimum amount of ``work'' required to convert the original distribution to the fitted one \cite{kantorovich_mathematical_1960, vaserstein1969markov}.
Section \ref{subsec:fittingexample} shows that, in many real-world cases, adopting this procedure is a sensible choice.

First, let us define the condition-compliant function given by

\begin{equation}
    y(x) =     
    \begin{cases}
    A \ \text{for} \ x < k, \\
    A e^{-B(x-k)^C} \ \text{for} \ x \geq k
    \end{cases}
    \label{eqn:curvefit}
\end{equation}
where $A, B, C, k$ are free parameters.
Consider only the points $x \in \{1, \dots, n \}$ such that the values obtained correspond to the proportion of individuals within each age group in the distribution.
The value $k$ is some integer that demarcates between the younger age groups assumed to contain equal proportions of individuals, and the older age groups where the proportion of individuals within a group decreases with the age of that group.
For example, $k=5$ would indicate that the \nth{5} age group will be where we shift between the cases in \autoref{eqn:curvefit}.

Values for $A, B, C$ are determined by fitting \autoref{eqn:curvefit} to the observed age distribution using a non-linear least squares method for each possible $k$-value, where $k \in \{1, \dots, n\}$. 
The best set of fitted parameters is determined based on which $k$-value (and associated values for $A, B, C$) minimises the value of the Wasserstein metric. 
Selecting the parameter set which leads to the lowest value of the Wasserstein metric ensures that our fitted age distribution departs from the original distribution as little as possible.

\begin{figure}[ht]
	\centering
	\includegraphics[width=0.6\textwidth]{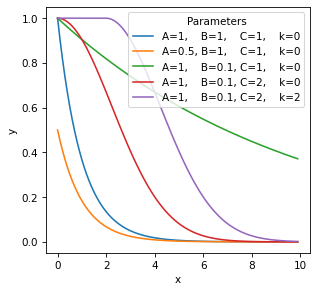}
	\caption{\textbf{Parameterisations of \autoref{eqn:curvefit}}. Various choices of parameter values chosen to demonstrate the flexibility of the functional form.}
	\label{figure:curve_demo}
\end{figure}

\section{Applications}\label{sec:examples}

We demonstrate the features of our modelling framework by using age distributions with 5-year age groups (0-4, 5-9, \dots, 95-99, 100+) at the ``Country/Area" (hereafter referred to as country) level for the year 2019 \cite{unpd_world_2019}. 
It is important to note that these broad age groups are chosen based on the granularity of the data, and do not reflect a requirement of the model, which admits age groups of any ``width''.
We show different cases using several countries.
First, we begin with the the stochastic process without activation rates (model 1).
Then we demonstrate use of the model with activation rates (model 2).
Finally, we use the curve-fitting approach combined with model 1.
In all cases, we run simulations for 350 timesteps (such that they reach the steady state) with a population of 10,000 agents.

\subsection{Using model 1}

\begin{figure}[ht]
	\centering
	\includegraphics[width=0.67\textwidth]{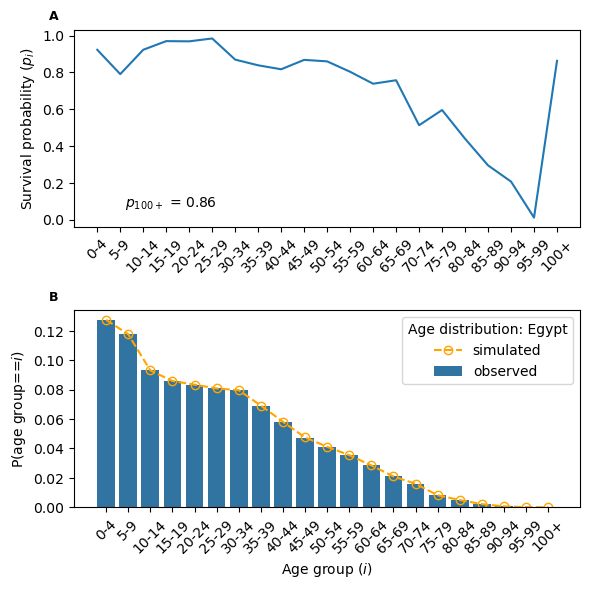}
	\caption{\textbf{Application of model to Egypt's age distribution.} A) indicates survival probabilities ($p_i$-values) calculated for Egypt, using a randomly selected $p_{100+}$ value which satisfies the conditions on $p_n$ for using the model without activation rates and B) shows Egypt's observed age distribution, and the age distribution generated from a simulation parameterised with the values from panel A).}
	\label{figure:demo_egypt}
\end{figure}

Egypt is an example of a population with monotonically decreasing age group sizes. 
In other words, it exhibits the stylised fact of a population pyramid with its base found in the youngest cohort.
Of the 201 countries included within our dataset, 57 (roughly 28\%) exhibit age distributions of this type.
These age distributions fulfil the parameter conditions set by model 1, so we can use this stochastic process.
This also means that, by solving the implied system of this stochastic process, we can directly obtain the survival probabilities needed to generate such age structures.
\autoref{figure:demo_egypt} demonstrates this by showing the age group survival probabilities and a perfect match between Egypt's empirical distribution and the steady-state age distribution produced by the model.

\subsection{Using model 2}

\begin{figure}[ht]
	\centering
	\includegraphics[width=0.67\textwidth]{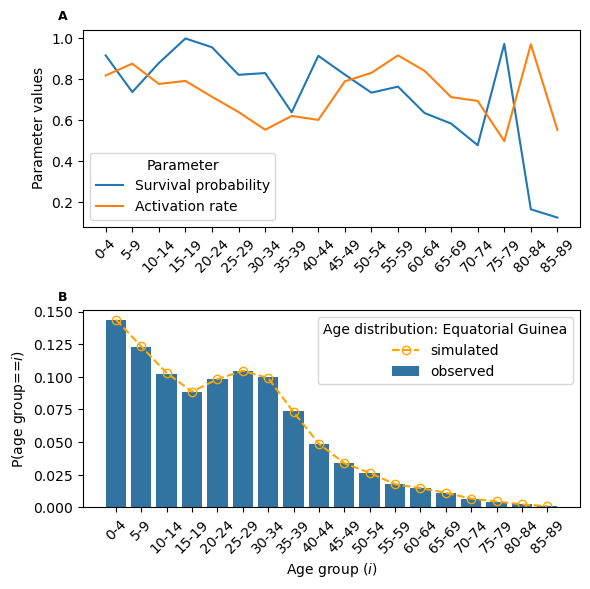}
	\caption{\textbf{Application of model with activation rates to Equatorial Guinea's age distribution.} A) indicates survival probabilities and activation rates, obtained using our optimisation method and B) shows Equatorial Guinea's observed age distribution, and the age distribution generated from a simulation parameterised with the values from panel A).}
	\label{figure:demo_eg}
\end{figure}

In Equatorial Guinea the population's age structure does not exhibit the pyramid stylised fact, so group size does not decrease monotonically with age.
This means that we cannot use model 1, as its parameter conditions are not met.
Instead, we use the model with activation rates (\autoref{figure:demo_eg}) as it allows more degrees of freedom.\footnote{As this population does not contain any individuals aged 90 and above, these age groups are dropped.}
\autoref{figure:demo_eg} shows, in the top panel, the age group survival probabilities and activation rates needed to generate Equatorial Guinea's age distribution.
The bottom panel demonstrates that, by using model 2, we get a highly accurate match between the the steady state age distribution and the empirical one with its characteristic hump in the 20-to-30 age groups.

Of the 144 countries that do not exhibit monotonically decreasing age distributions (where model 1 is ideal), we find that 45 can be successfully simulated using model 2. 
We use as the criteria for success that the mean absolute error between the simulated steady state age distribution and the observed age distributions is below a threshold of $1 \times 10^{-4}$. 
If we cannot arrive at a combination of survival and activation rates that achieve this level of accuracy for a given country within 250 iterations of the optimisation method, we assume that model 2 is unsuitable for simulating that country's age distribution. 
For countries falling into this group, we use the curve-fitting procedure.

\subsection{Curve-fitting procedure}\label{subsec:fittingexample}

The United Kingdom (UK) is representative of the age structure of several European nations with a large ageing population.
Here, we observe that the largest groups are in their 30s and 50s.
The youngest groups are smaller, but not in a drastic way, giving the distribution a flat structure across the ages 0 to 59 years.
By looking at this stylised shape, it is clear that fitting \autoref{eqn:curvefit} to the data is a sensible choice. 

\begin{figure}[ht]
	\centering
	\includegraphics[width=0.67\textwidth]{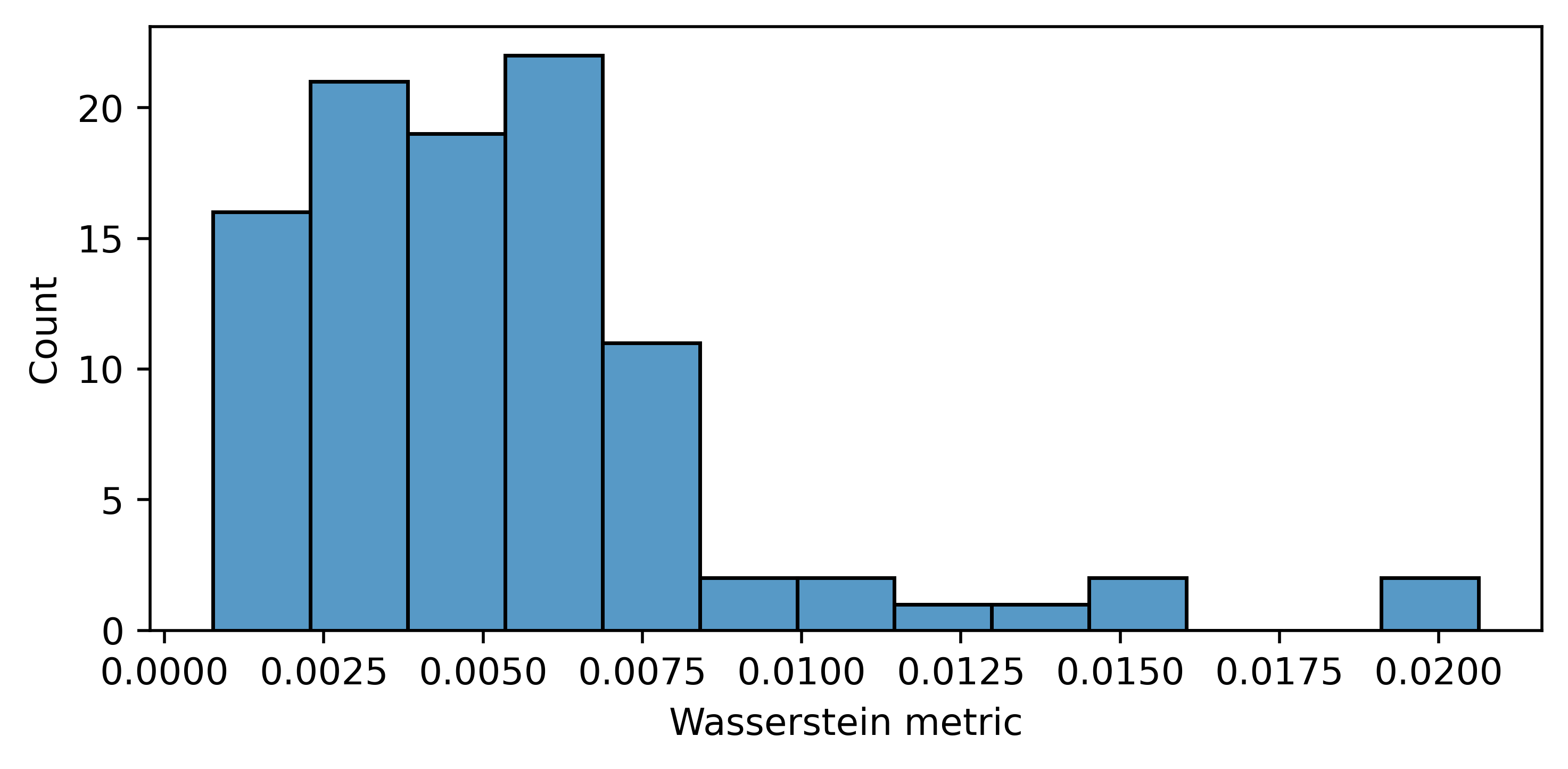}
	\caption{\textbf{Wasserstein metric values calculated based on the ``work'' needed to convert an observed age distribution to the corresponding fitted age distribution.} The mean value of the Wasserstein metric, across the 99 countries for which we perform the curve fitting, is approximately $0.0055$.}
	\label{figure:wassmetric}
\end{figure}

\begin{figure}[ht]
	\centering
	\includegraphics[width=0.67\textwidth]{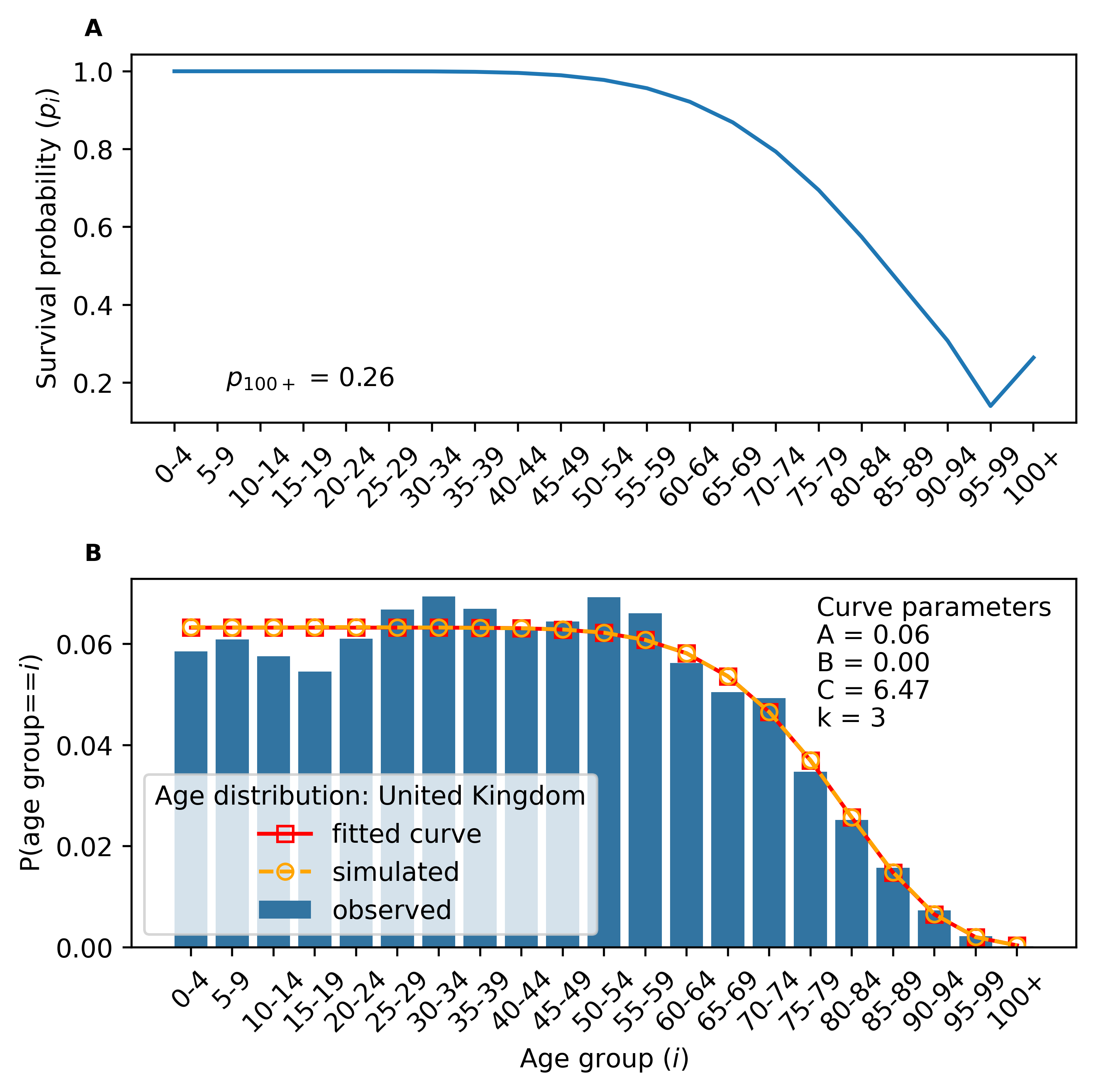}
	\caption{\textbf{Application of model to a fitted version of the United Kingdom's age distribution.} A) indicates survival probabilities ($p_i$-values) calculated for the fitted version of the UK age distribution, as shown in B), using a randomly selected $p_{100+}$ value which satisfies the conditions on $p_n$ for using the model without activation rates and B) shows the United Kingdom's observed age distribution, the curve fit to the age distribution (with associated parameters), and the age distribution generated from a simulation parameterised with the values from A) to reproduce the fitted age distribution.}
	\label{figure:demo_uk}
\end{figure}

The UK is one of 99 countries for which neither model 1 nor model 2 prove suitable, and for which we apply the curve fitting procedure. 
The distribution of the Wasserstein metric values for these 99 countries is shown in \autoref{figure:wassmetric}, with 56 countries having below-average metric values. 
The UK is one of these 56 countries, with a Wasserstein metric value of approximately $0.0027$. 
This indicates that the fitted UK age distribution is quite similar to the observed distribution (in the context of the group of age distributions we consider).
This further affirms our feeling that it would be sensible to apply the fitting method to the UK's age distribution.
Thus, we fit and use model 1 to obtain the survival probabilities that allow the stochastic process to generate the fitted distribution.
Panel A in \autoref{figure:demo_uk} shows the estimated survival probabilities.
Panel B shows the empirical UK age distribution, the fitted one, and the steady-state one generated by model 1.

In summary, the proposed method is able to provide the parameters needed to generate a large number of empirical age distributions for agent-based models fulfilling the characteristics enlisted in the introduction of this paper.
Our examples show its versatility using country-level real-world data.
Next, we close the paper with a summary and discussion on our findings and the contribution of the method.

\section{Discussion and conclusions}

\subsection{Limitations}

The main limitation of the proposed method is that it only applies to ABMs with steady-state population dynamics with exogenous survival probabilities.
This means that, in systems where the probability of surviving is affected by an endogenous variable--such as an infectious disease--it would not be possible to calculate the survival rates needed to generate a target age distribution.
Potential future extensions could consider more elaborate stochastic processes that incorporate such variables.

Another limitation is that we assume a constant population size in the ABM that needs to be calibrated.
Extending the model to a growing population introduces some complications such as the loss of conservation restrictions in the system of equations that needs to be solved.
Non-conservative models are common in population ecology, where researchers tends to focus on using observed rates (of survival and of fecundity) to make predictions about the steady state distribution of ages/life stages. 
In many social systems, we have the opposite problem: given a known steady-state age/life distribution, we need to determine rates (of survival and of activation) that will lead to the desired steady-state distribution.
In fact, models 1 and 2 are the social analogues to the popular Leslie \cite{leslie_use_1945,leslie_further_1948} and Lefkovitch \cite{lefkovitch_study_1965} matrices found in population ecology. 
Another potential direction of future research is to extend our models to consider non-conservative populations and to uncover a more direct connection with the population ecology literature.

\subsection{On the evolution of the age distribution}

A potential concern with our method is that the dynamics of age distributions are not considered.
We believe that this is not a fundamental problem, as there are ways in which steady-state transitions could be induced through a careful specification of the survival probabilities across time.
In fact, a common practice in the ABM and micro-simulation communities is to initialise a model with the empirical age structure and survival probabilities, let is run, and interpret the final age distribution demographic forecast.
We advise against this practice as, often, initialising an ABM with empirical data does not guarantee that the system has stabilised in a state that would be representative of the real world.
Quite the opposite, a data-induced initialisation may generate transient dynamics that do not reflect the natural evolution of the system; i.e. it would only yield a calibration artefact.
Instead, we recommend to first obtain stable dynamics such as those of a steady state.
In the context of demographic change, calibrating for steady-state dynamics is a reasonable approach.
For instance, consider conducting a census to construct the age distribution of a population.
If one would repeat this exercise a year later, it would be unlikely to observe a radical shift in the age structure.
Such shifts tend to be generational, and are usually considered to be of a long-term nature.
Thus, by initialising an ABM in a steady state, one is simply assuming we are capturing some of the short-term dynamics taking place in the real world.
Then, we can proceed to perturb the system to analyse long-term changes.

One particular type of perturbation is precisely the inter-temporal change of survival probabilities.
For example, suppose that one is interested in studying the labour-market impacts of an age structure transition from a pyramid-shape distribution to an inverted-pyramid one (the latter perhaps predicted by a demographic model).
If one would know the initial and final shapes (or even the intermediate), then our method can be applied to obtain the survival probabilities that yield such distributions in the steady state.
These parameters can be progressively imputed in the model as it moves from one steady state to another, just as demographics change marginally in short periods.
This strategy would reduce potential biases from the transient dynamics introduced by a purely empirical initialisation, and provide a more reliable way to study the nuanced impacts of a changing age distribution.

\subsection{Conclusions}

This paper introduces a method for generating steady-state age distributions in agent-based models. 
We highlight the flexibility of this method by illustrating its use on several country-level age distributions differing in shape.
These results demonstrate the ease with which our method can be implemented across a broad range of scenarios.

Users should consider the type of age distribution they are interested in when deciding which of the three approaches to employ, as their efficacy depends on the relative sizes of the age groups within the population.
As a rule of thumb for determining which model to try first, we suggest the following.
The model without activation rates (model 1) is suitable for modelling populations with expansive population pyramids. 
The model with activation rates (model 2) may be useful for modelling populations with stationary or constrictive population pyramids (with a hump).
Finally, using a combination of curve-fitting and model 1 may be successful for some populations with stationary population pyramids.

\bibliography{AgeDistributions}

\section*{Acknowledgements}

\noindent We are thankful to Robert Axtell for his insightful comments on the links between this work and population ecology.
We are also grateful to Mark Birkin for his micro-simulation insights.

\noindent This work was supported by Wave 1 of The UKRI Strategic Priorities Fund under the EPSRC Grant EP/W006022/1, particularly the ``Shocks and Resilience'' theme within that grant \& The Alan Turing Institute.

\section*{Code availability}
\noindent Code used for data analysis, parameter fitting, and simulation is available in a GitHub repository (\url{https://github.com/k3fair/agedist-gen}).

\end{document}